\documentclass[12pt]{article}
\usepackage[british]{babel}
\usepackage{graphicx}
\usepackage{txfonts}

\newcommand{\oii}{[OII]}

\newcommand{\nii}{[NII]}
\newcommand{\ha}{H$\alpha$}
\newcommand{\hb}{H$\beta$}
\newcommand{\msy}{$\rm{M_{\odot}\,yr^{-1}}$}

\newcommand\apj{ApJ}
\newcommand\aap{A\&A}

\newcommand\mnras{MNRAS}

\newcommand\apjl{ApJL}

\newcommand\aj{AJ}
\newcommand\pasp{PASP}

\newcommand\pubnumber{Article 2 in eConf C1304143}
\newcommand\pubdate{\today}

\def\napoli{$^{1}$MPE;
$^{2}$DARK;
$^{3}$IAA-CSIC;
$^{4}$The University of Leicester;
$^{5}$TLS Tautenburg;
$^{6}$Czech Technical University in Prague;
$^{7}$Universit\`a degli studi di Milano-Bicocca;
$^{8}$CfA;
$^{9}$Pennsylvania State University;
$^{10}$University of Warwick;
$^{11}$UPV/EHU-IAA/CSIC;
$^{12}$Ikerbasque, Basque Foundation for Science
}

\def\support{\footnote{Based on observations made with telescopes at the European Southern Observatory at La Silla/Paranal, Chile under program 090.A-0760(A).}}

\def\Title#1{\begin{center} {\Large #1 } \end{center}}
\def\Author#1{\begin{center}{ \sc #1} \end{center}}
\def\Address#1{\begin{center}{ \it #1} \end{center}}

\newcommand\pubblock{\rightline{\begin{tabular}{l} \pubnumber\\
         \pubdate  \end{tabular}}}
\newenvironment{Abstract}{\begin{quotation}  }{\end{quotation}}
\newenvironment{Presented}{\begin{quotation} \begin{center} 
             PRESENTED AT\end{center}\bigskip 
      \begin{center}\begin{large}}{\end{large}\end{center} \end{quotation}}

\def\beq{\begin{equation}}
\def\eeq#1{\label{#1}\end{equation}}
\def\eeqn{\end{equation}}

\def\beqa{\begin{eqnarray}}
\def\eeqa#1{\label{#1}\end{eqnarray}}
\def\eeqan{\end{eqnarray}}

\let\bar=\overbar

\def\Dslash{\not{\hbox{\kern-4pt $D$}}}
\def\dslash{\not{\hbox{\kern-2pt $\del$}}}

\def\msb{{\bar{\ssstyle M \kern -1pt S}}}

\begin{document}
\begin{titlepage}
\pubblock

\renewcommand*{\thefootnote}{\fnsymbol{footnote}}
\vfill
\Title{The low-extinction afterglow in the solar-metallicity host galaxy of $\gamma$-ray burst 110918A\support}
\vfill
\Author{J. Elliott$^{1}$, T.~Kr{\"u}hler$^{2}$, J.~Greiner$^{1}$, S.~Savaglio$^{1}$, F.~Olivares~E.$^{1}$, A.~Rau$^{1}$, A.~de~Ugarte~Postigo$^{3,2}$, R.~S{\'a}nchez-Ram{\'i}rez$^{3}$, K.~Wiersema$^{4}$, P.~Schady$^{1}$, D.~A.~Kann$^{5,1}$, R.~Filgas$^{6,1}$, M.~Nardini$^{7}$, E.~Berger$^{8}$, D.~Fox$^{9}$, J.~Gorosabel$^{3,11,12}$, S.~Klose$^{5}$, A.~Levan$^{10}$, A.~Nicuesa Guelbenzu$^{5}$, A.~Rossi$^{5}$, S.~Schmidl$^{5}$, V.~Sudilovsky$^{1}$, N.~R.~Tanvir$^{4}$, C.~C.~Th{\"o}ne$^{3}$}
\Address{\napoli}
\vfill
\begin{Abstract}
Galaxies selected through long $\gamma$-ray bursts (GRBs) could be of fundamental importance when mapping the star formation history out to the highest redshifts. Before using them as efficient tools in the early Universe, however, the environmental factors that govern the formation of GRBs need to be understood. Metallicity is theoretically thought to be a fundamental driver in GRB explosions and energetics, but is still, even after more than a decade of extensive studies, not fully understood. This is largely related to two phenomena: a dust-extinction bias, that prevented high-mass and thus likely high-metallicity GRB hosts to be detected in the first place, and a lack of efficient instrumentation, that limited spectroscopic studies including metallicity measurements to the low-redshift end of the GRB host population. The subject of this work is the very energetic GRB 110918A ($E_{\gamma,\mathrm{iso}}=1.9\times10^{54}\,\mathrm{erg}$), for which we measure one of the largest host-integrated metallicities, ever, and the highest stellar mass for $z<1.9$. This presents one of the very few robust metallicity measurements of GRB hosts at $z \sim 1$, and establishes that GRB hosts at $z \sim1$ can also be very metal rich. It conclusively rules out a metallicity cut-off in GRB host galaxies and argues against an anti-correlation between metallicity and energy release in GRBs.

\end{Abstract}
\vfill
\begin{Presented}
the Seventh Huntsville Gamma-Ray Burst Symposium\\
Nashville, Tennessee, 14--18 April 2013
\end{Presented}
\vfill
\end{titlepage}
\def\thefootnote{\fnsymbol{footnote}}
\setcounter{footnote}{0}

\section{Introduction}
During their prompt emission, long gamma-ray bursts (GRBs) are the brightest objects in the Universe, easily reaching isotropic-equivalent luminosities as high as $\sim10^{54}\,\mathrm{erg\, s^{-1}}$. Their observed association to supernovae events has tightly linked them to the death of massive stars~\cite{Hjorth12a}. The GRB progenitors are likely Wolf-Rayet like stars, that usually undergo vigorous mass loss from stellar winds, and so metallicity constraints ($Z<0.3\,\mathrm{Z_{\odot}}$) on the progenitor are postulated to ensure a GRB occurs~\cite{Hirschi05a, Yoon05a, Woosley06a}.

The possible association of long GRBs with massive stars supported the idea that they could be used as complementary and independent tracers of star formation, due to their very high luminosities~\cite{Daigne06a,Li08a}. However, to have full confidence in these studies the intrinsic evolutionary effects in long GRB production must be understood and the galactic environments preferred by the progenitor need to be quantified~\cite{Salvaterra12a,Robertson12a,Elliott12a}. Of particular interest is the relation between the galaxies selected by GRBs and the star formation weighted population of field galaxies. To be direct and unbiased tracers of star formation, the relative rates of GRBs in galaxies of various physical properties should be the same in galaxies taken from samples that trace the global star formation density at a given redshift. Studies based on these galaxy samples are most commonly performed at $z \lesssim 1.5$, where the star formation of field galaxies is largely recovered by state-of-the-art deep-field surveys.

Only a few host galaxies with substantial gas-phase metallicities around or above solar~\cite{Levesque10b} that directly violate the proposed cut-off in galaxy metallicity have been observed to date. There is thus still lively debate in the literature about the nature of GRB hosts, and their relation to the star formation weighted galaxy population as a whole \cite{Niino11a, Mannucci11a, Kocevski11a, Graham12a}. GRB hosts with high stellar mass and high global metallicity are hence of primary interest for GRB host studies, as they directly probe this allegedly forbidden parameter space.

\section{Observations}

\begin{figure}[htb]
\centering
\includegraphics[width=9cm]{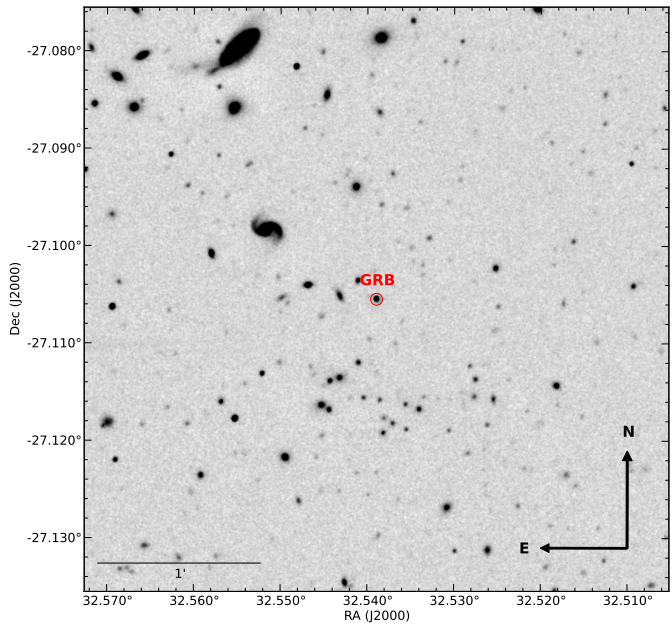}
\caption{A GROND $i'$ band image showing the field of GRB 110918A, and the position of the afterglow highlighted by the red circle.}
\label{fig:fits_grb}
\end{figure}

The luminous GRB 110918A was detected on the 18th of September 2011 at $T_{0}$=21:26:57 UT \cite{Hurley11a}. This burst had one of the highest fluences of any GRB observed over the last 20 years and had the highest peak flux ever detected by {\it Konus-Wind} \cite{Golenetskii11a,Frederiks11a}. The X-ray and optical afterglows (see Fig.~\ref{fig:fits_grb}) were later found~\cite{Krimm11a,Mangano11a}, and followed, by several instruments, for which we outline the ones relevant to our work in the following.

The Gamma-Ray Burst Optical Near-infrared Detector~(GROND; \cite{Greiner08a}) mounted at the MPG/ESO $2.2~\mathrm{m}$ telescope at La Silla, Chile, began its follow-up campaign of GRB 110918A $29.2~\mathrm{hrs}$ after the trigger simultaneously in the $g^{\prime}r^{\prime}i^{\prime}z^{\prime}JHK_{s}$ filters \cite{Elliott11a}. Spectra of the afterglow were obtained with the Optical System for Imaging and low Resolution Integrated Spectroscopy~(OSIRIS; \cite{Cepa00a}) mounted at the Gran Telescopio Canarias (Roque de los Muchachos) and the Gemini Multi-Object Spectrographs~(GMOS; \cite{Hook04a})  mounted at the Gemini North telescope (Mauna Kea) at $\sim1$ day after the trigger, for which a redshift of $z=0.984\pm0.001$ was estimated using the transition of several metal ions including Fe~II, Mg~II, and Mg~I.

GROND observations continued for over 1 month after the GRB trigger and an underlying host was discovered. Further deep observations of the host galaxy were made 392 d after the trigger with the Wide Field Imager~(WFI; \cite{Baade99a}), also mounted on the MPG/ESO 2.2m telescope and a source at the host's location found in the Wide-field Infrared Survey Explorer~(WISE; \cite{Wright10a}) All-Sky Source Catalogue. Finally, late time spectra were obtained for the underlying host galaxy with OSIRIS ($\sim40$ days after the burst) and with X-shooter~\cite{Vernet11a} mounted at the Very Large Telescope (Kueyen/UT2), approximately 1 year later.

\section{Results}
\subsection{Afterglow}
A broadband SED was constructed from the optical/NIR GROND photometry $\sim2$ days after the burst and fit with a powerlaw, assuming that the afterglow emission is well described by the standard synchrotron mechanism. Most noteable from the best fit, is the line of sight extinction, $A^{\mathrm{GRB}}_V=0.16\pm0.06\,\mathrm{mag}$, that assumes a Small Magellanic Cloud (SMC) like dust.

\subsection{Host Galaxy}
\begin{figure}[htb]
\centering
\includegraphics[trim=0.5cm 0cm 0cm 0cm, clip=True, width=9.0cm]{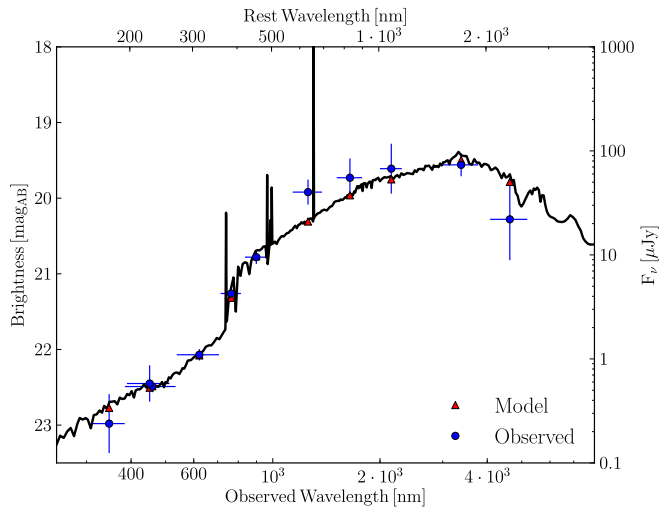}
\caption{The SED of the host of GRB 110918A obtained using GROND, WFI and WISE data, amounting to 11 filters: $UBg'r'i'z'JHK_{s}W1W2$ from left to right. The best-fit spectrum is depicted in black.}
\label{fig:_sedhost}
\end{figure}

The host of GRB 110918A was detected in 11 different filters ranging from the ultra-violet to $4.5\mu$m, yielding a well-sampled photometric SED (see Fig.~\ref{fig:_sedhost}). To estimate the global properties of the host galaxy we employed standard techniques that use stellar population synthesis to estimate stellar masses, the full details can be found in~\cite{Elliott13a}. This reults in a stellar mass of $\log_{10}\left(\mathrm{\frac{M_{*}}{M_{\odot}}}\right)=10.68\pm0.16$.

From the X-shooter/OSIRIS spectra, we clearly detect the H$\alpha$ and H$\beta$ transition from the Balmer series, as well as the forbidden transitions of \oii($\lambda\lambda$3726, 3729) and \nii($\lambda$6584). The Balmer ratio of \ha/\hb~implies an $E(B-V)^{\rm{gas}}=0.57^{+0.24}_{-0.22}$ mag towards the star forming regions, assuming a Milky-Way like extinction law. Utilising the \ha~we estimate a star formation rate of SFR$_{H\alpha}=40$~\msy. Using the different emission line ratios, we measure a metallicity of 12 + log(O/H)$_{\rm{N2H\alpha}} = 8.93 \pm 0.13$ and 12 + log(O/H)$_{\rm{N2O2}} = 8.85_{-0.18}^{+0.14}$ using the formulation of \cite{Nagao06a}. Different calibrations of the strong-line diagnostics yield consistent values and imply metallicities between 0.9 and 1.7 times solar. 

\section{Discussion}

\subsection{The Afterglow \& Host Dynamic}

\begin{figure}[htb]
\includegraphics[trim=0cm 0cm 0cm 0cm, clip=True, width=7.8cm]{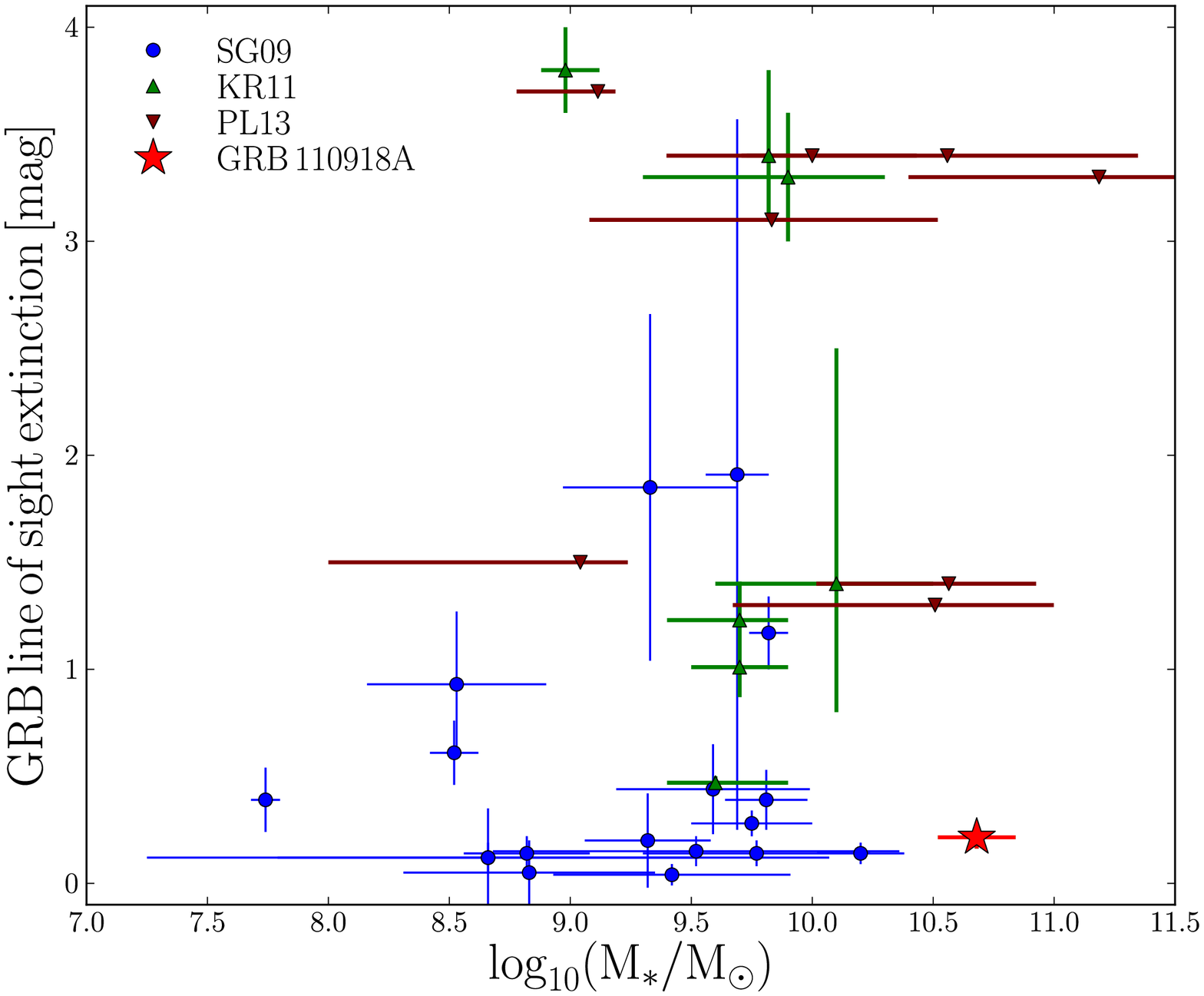}\includegraphics[width=8.3cm]{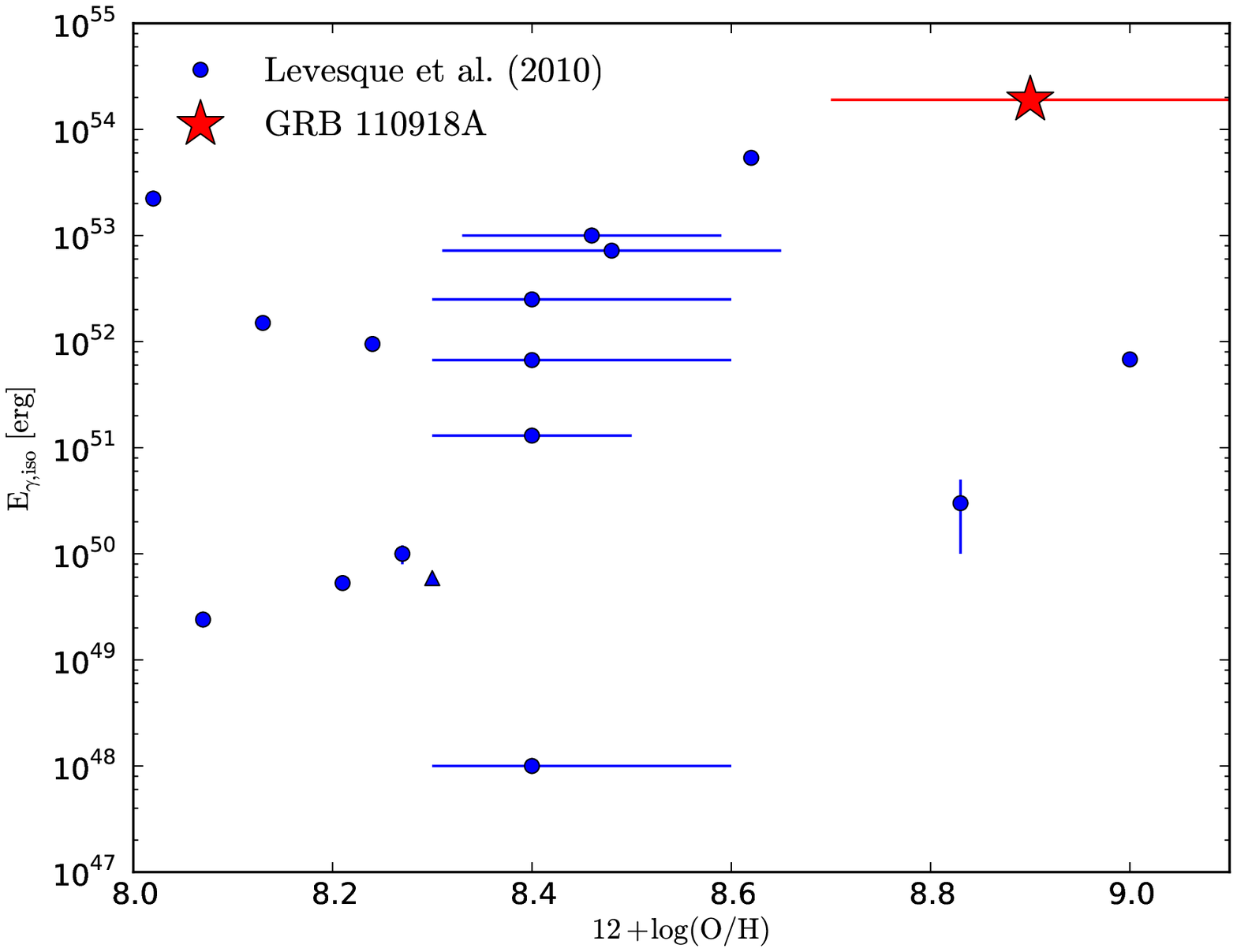}
\caption{{\bf Left}: The GRB line of sight $A^{\mathrm{GRB}}_{V}$ plotted against the stellar mass of the host galaxy. The extinction values have been obtained from~\cite{Kann06a},~\cite{Kann10a},~\cite{Greiner11a},~\cite{Schady12a} and~\cite{Perley13a}. {\bf Right}: The isotropic-equivalent energy release in $\gamma$-rays of GRBs plotted against the gas-phase metallicity of the host galaxy. Blue data are taken from~\cite{Levesque10a}. }
\label{fig:grbav}
\end{figure}

There is a very tight correlation between stellar-mass and sight-line extinction probed by the GRBs~\cite{Perley13a}, and is found to be stronger than for any other physical property of the galaxy, such that hosts selected due to high afterglow extinction have systematically more massive and dust extinguished sight lines than the optically selected hosts (Fig.~\ref{fig:grbav}). While in principle, cases like GRB 110918A would be easy to identify (bright afterglow, easy localization, bright host), no comparable example has been reported in the literature to date. It is possible that: (i) the geometry of dust within the host of GRB110918A is more patchy than homogeneous (see also, GRB 100621A and 061222A~\cite{Kruehler11a,Perley13a}, or (ii) the progenitor had enough time to destroy local dust from its UV emission~\cite{Perley13a}.

\subsection{Fundamental Metallicity Relation}

The difference between galaxies of long GRBs and that of normal star forming field galaxies is still an on going debate. We have derived estimates for the mass, metallicity and SFR of the host of GRB 110918A, which facilitates comparing this galaxy with respect to normal star forming galaxies through the fundamental metallicity relation~\cite{Mannucci11a}. The FMR determined value is in agreement with the metallicity from the \nii/\ha~line ratio. This illustrates that the mass and SFR of a GRB-selected galaxy, at least for this one event, can be used as a fair proxy for the metallicity even in the solar, or super-solar regime. 

\subsection{Metallicity and Long GRB Progenitors}
Many authors have attributed the fact that most long GRB host galaxies exhibit low metallicities as the result of an environmental preference~\cite{Modjaz08a,Graham12a,Perley13a}. This dependence on metallicity has also led to the prediction that the lower the progenitor metallicity, the larger the angular momentum, and thus the higher the energy output of the GRB~\cite{MacFadyen99a}. With one of the largest metallicities and isotropic-equivalent prompt energies, ever, GRB 110918A is in contradiction to the idea of a correlation between $E_{\gamma,\mathrm{iso}}$ and metallicity (Fig.~\ref{fig:grbav}), consistent with other studies~\cite{Wolf07a,Levesque10a}.

\subsection{Metallicity Cut-off}
Recently, \cite{Perley13a} performed an extensive photometric study of host galaxies selected from a sample of dark bursts, limiting the selection biases present in previous works. However, while the inclusion of dark GRB hosts increases the consistency of GRB hosts with the star formation weighted sample of field galaxies, there is still a clear lack of high-mass galaxies at $z \lesssim 1.5$. Associating the galaxy mass with metallicity, this provides indirect evidence for a metallicity effect in GRB hosts. A similar conclusion was reached based on a comparison of long GRB hosts with supernovae hosts~\cite{Graham12a}, namely that long GRB hosts show a strong preference for lower metallicity environments relative to other populations of star forming galaxies, with a metallicity cut-off of $Z<0.5\,\mathrm{Z_{\odot}}$. This cut-off is not consistent with the host galaxy of GRB 110918A even if dispersions of 0.3 dex are considered.

\section{Summary}
A dedicated follow-up campaign of the afterglow of GRB 110918A and its underlying host has revealed a unique system with the following properties: (i) one of the most massive galaxies selected by a GRB at $z\sim1$, (ii) the first relatively unobscured afterglow that has been detected in a very massive and dusty host galaxy, (iii) one of the most metal-rich long GRB host galaxies found yet, (iv) no different to star forming galaxies selected through their own stellar light, when compared using the FMR, (v) opposes an anti-correlation between energy output of the GRB and environmental metallicity, and (vi) contradicts a cut-off for host galaxies of $Z<0.5\,\mathrm{Z_{\odot}}$.



\end{document}